# Privocracy: Online Democracy through Private Voting

Pedro Camponês
p.campones@campus.fct.unl.pt
NOVA LINCS & NOVA School of
Science and Technology
Lisbon, Portugal

Hugo Pereira
hg.pereira@campus.fct.unl.pt
NOVA LINCS & NOVA School of
Science and Technology
Lisbon, Portugal

Adrian Persaud
apersau@purdue.edu
Purdue University
West Lafayette, IN

Kevin Gallagher
k.gallagher@fct.unl.pt
NOVA LINCS & NOVA School of
Science and Technology
Lisbon, Portugal

Santiago Torres-Arias
santiagotorres@purdue.edu
Purdue University
West Lafayette, IN

**Abstract**

In traditional access control policies, every access granted and administrative account introduces an additional vulnerability, as a corruption of a high-privilege user can compromise several sensitive files. *Privocracy* is an access control mechanism that minimizes the need to attribute high privileges by triggering a secure e-voting procedure to run commands that require using sensitive resources. With Privocracy an organization can distribute trust in resource access, minimizing the system vulnerabilities from single points of failure, all while maintaining the high flexibility of discretionary access control policies.

The Privocracy voting mechanism achieves everlasting privacy, ensuring votes remain confidential regardless of an adversary's computational power, while addressing the dependability requirements of a practical and secure system. The procedure incorporates useful features such as vote delegation to reduce voter fatigue, rapid voting rounds to enable quick action during emergencies, and selective vote auditing for application-level accountability. Our experimental results demonstrate that Privocracy processes votes efficiently and can be deployed on commodity hardware.

## 1 Introduction

Access control is a process that safeguards a computer system's security by defining and enforcing policies governing which users may access which resources and under which conditions. These mechanisms protect files from unauthorized disclosure, modification, or execution. Access control policies inherently involve trade-offs between security and usability [63]: discretionary policies emphasise flexibility by allowing resource owners to grant permissions, while mandatory policies prioritize security through centrally enforced, label-based policies. Hybrid approaches between discretionary and mandatory policies [6, 19, 47, 64] balance these concerns, tailoring the trade-off to the needs specific systems.

Real-world deployments commonly introduce administrative hierarchies to manage permissions at scale. Enterprise systems, RBAC [64] configurations, and security frameworks such as Active Directory [62] rely on privileged administrators who assign or delegate permissions to less-privileged users. However, such administrative hierarchies introduce critical single points of failure: the compromise of a highly privileged account can enable the attacker to escalate privileges, grant unauthorized permissions, or modify sensitive configuration files. Although preventive measures, such as strong authentication, perimeter security, and monitoring, can reduce exposure, traditional access-control architectures lack built-in dependability mechanisms to limit the impact of a privileged-user compromise [4].

These risks are well documented in widely adopted security guidelines, such as the Open Web Application Security Project (OWASP) [58], which highlights *Broken Access Control* and *Security Misconfiguration* as recurring critical vulnerabilities, many stemming from mismanaged administrative roles. Even with carefully designed policies, system integrity often hinges on the correct behavior of a small number of highly authorized users, revealing a structural fragility that conventional models do not address.

To mitigate this class of vulnerabilities, we introduce *Privocracy*, a collective based access control system that removes unilateral control over sensitive operations. Instead of allowing a single party to execute critical actions, Privocracy requires approval from multiple independent entities through an e-voting process. The system's security thus relies on preventing adversaries from garnering enough voting power to reach the authorization threshold or from suppressing honest votes needed to achieve it. This distributed decision structure eliminates single points of failure and increases robustness to administrator compromise. In addition, Privocracy can be deployed atop existing discretionary access control systems with minimal modifications, keeping integration overhead low and requiring only a small learning curve for deployment in real systems. While ideas about distributed trust in access control procedures have been discussed [41], as far as we are aware, Privocracy is the concrete construction of a practical access control mechanism that distributes trust through a private e-voting procedure. All while maintaining the flexibility of DAC mechanisms.

Private e-voting algorithms have been extensively studied in the context of large public elections [12, 13], where achieving ballot privacy [14, 45] and end-to-end verifiability [31] is essential. These systems are typically designed for settings involving infrequent but large-scale electoral events [53], and therefore optimize for high voter throughput and strong auditability.

In contrast, an e-voting mechanism for access control must support rapid decision-making among a small group of participants, operate reliably over long-term deployments, and satisfy the specific requirements of authorization workflows. Such systems must remain responsive, fault-tolerant, and robust under repeated use, which distinguishes them from traditional election-oriented e-voting protocols.

In Privocracy, different voters can have different (dynamically configurable) weights in the decision of an election outcome, allow the delegation of votes among voters so the system can operate efficiently even when some voters are inactive, and enable fast voting rounds for the execution of emergency commands.

Within our specified system model (§ 2) and following established primitives (§ 3), our contributions are as follows:

- We specify (§ 4) an access control policy that allows the collaborative deliberation of user requests and show how to use Privocracy to define such a policy.
- We specify (§ 5) and implement (§ 6) Privocracy, an open-source access control system with distributed trust that enables private, efficient, and auditable e-voting.
- We evaluate the performance and measure the upper bound leakage introduced by Privocracy (§ 7). Our results show Privocracy acheives low end-to-end latency in elections (100 seconds for 60 voters), while being able to run on commodity hardware.

Finally, we explore related work (§ 8) in voting systems and access control systems, demonstrating why state-of-the-art solutions are inadequate to comply with our objectives.

## 2 System Model

**Architecture Overview:** The system is comprised by the *Operational space* and the *Voting space*, illustrated in figure 1. The Operational instances are components of the Operational space, while Voter instances comprise the Voting space.

Operational instances may either be replicas or have specific roles, such as e-mail or file servers. These instances have resources whose access is mediated through the Privocracy access control mechanism, which extends the default Discretionary Access Control (DAC) mechanism in Linux. With DAC policies, users have limited permissions to the operational instances' resources, but can be upgraded to unrestricted access to sensitive files. In Privocracy, an operation attempting to access restricted resources triggers a voting procedure to determine whether the operation executes. For non-sensitive files, the established rules of the underlying DAC stay in effect.

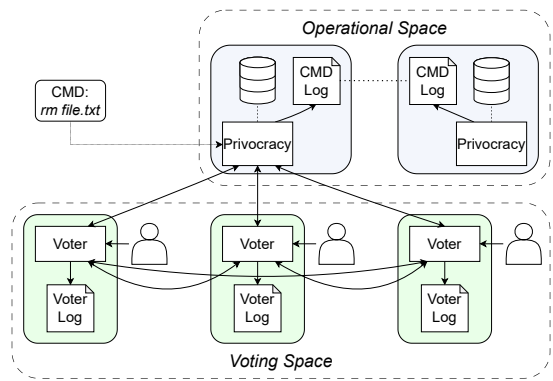

**Figure 1. Architecture Overview:** An operational instance hosts a Privocracy frontend, a Command Log, and serves a particular role (e.g., mail server, file server, etc.). Users issue commands to the Privocracy frontend, which disseminates it to Voter instances. Each Voter will then require manual intervention to vote on whether the operation should be executed. During the voting procedure the Voter instances will communicate with each other in order to blind the individual votes and compute partial vote tallies. Upon concluding the voting procedure, the Voters send their partial tallies to Privocracy, for it to compute the final vote tally and execute the operation if the tally exceeds a configurable threshold.

A command requests access to a protected resource by triggering the Privocracy frontend module which will receive the command, trigger an election to determine whether it should be executed, and execute it or not depending on the voting results. In this process, the request is forwarded to the Voter instances and each of which will require manual intervention to vote on whether the command is executed.

**Adversary Model:** We assume two distinct fault models. The Operational space operates under an honest-but-curious fault model, while the Voting space assumes byzantine faults.

The Operational space adversary can perform passive attacks in all processes. To protect sensitive operations, such as reconstructing the tally and checking whether the approval threshold is met, we assume these computations occur inside a Trusted Execution Environment (TEE), which ensures confidentiality of the election result. However, if a deployment cannot rely on TEEs, we discuss the resulting leakage channels and their security implications in § 5.5. Additionally, we assume processes can crash and be recovered at any point in the protocol, but there must be at least one correct instance in execution throughout the protocol.

In the Voting space we follow a byzantine fault model [52] comprised of $n$ processes, where the adversary may corrupt up to $f < \frac{n}{3}$. A corrupted process is considered *faulty* and behaves arbitrarily; otherwise, it is *correct*.

Processes can verify message authenticity, messages cannot be lost, altered, or duplicated in transit, and the adversary cannot forge messages from correct processes. We consider adversarial corruption as anything that hinders a process from executing as intended. In practice, these corruptions can result from either deliberate attacks by an adversarial entity or hardware/software faults.

Finally, the adversary can delay and reorder messages arbitrarily, determining the scheduling of message delivery. However, they cannot prevent messages from eventually reaching their intended recipient. A notable impact of this assumption is that, when the result of an algorithm depends on the order in which the system processes messages, the adversary can determine its output.

**Cryptographic Assumptions:** In the Operational space we consider an *unconditional security* setting where the adversary's computational power is unbounded. While in the Voting space we assume a *computational security* setting, where the adversary's computational power is polynomially bounded in a security parameter $\lambda$.

**Network Model:** This work follows a partially-synchronous network model [38], where the system initially behaves asynchronously but eventually reaches a *Global Stabilization Time* (GST), after which there exists a known upper-bound $\delta$ on message delivery time. We do not consider omission faults, such as those that might occur during a network partition. In these scenarios, we assume that processes will reconnect after the partition ends and retransmit the original messages; we attribute the long delay in such messages in transit to the inherent asynchrony of the Internet.

## 3 Background

Privocracy extends the traditional discretionary access control of Unix systems by issuing an e-voting procedure to decide whether a command should be executed. This voting procedure must both blind individual user votes and yet allow these votes to be auditable. Furthermore, this system should have high availability and tolerate adverse network conditions. To achieve these requirements, the voting algorithm uses Secret Sharing based Multiparty-Computation techniques to correctly compute the vote tallies, alongside with Zero-Knowledge proofs to ensure the integrity of votes.

**Discretionary Access Control (DAC):** Linux systems implement a discretionary access control model in which resource owners determine how their files may be accessed. Every file is associated with an *owner*, a *group*, and a set of permission bits specifying read ($r$), write ($w$), and execute ($x$) privileges for the owner, the group, and all other users. The owner of a file may modify these permissions at will, thereby delegating access to other users or to groups, through the use of Access Control Lists. System administrators manage the creation of users and groups, while file owners select which group a file belongs to. Processes execute with the user and group identifiers of the invoking principal, and thus inherit the permissions granted to those identities.

**Secret Sharing** schemes allow a dealer to divide a secret into multiple *shares* such that any individual share reveals no information about the secret, and the secret can be recovered only by combining a subset of shares [15, 65]. A scheme consists of a *sharing* phase, where the dealer distributes the shares, and a *reconstruction* phase, where parties combine their shares to recover the secret.

**Definition 3.1.** A Secret Sharing (SS) protocol satisfies:

- **Secrecy**: If the dealer is honest, no information about the secret is leaked prior to reconstruction.
- **Termination**: If the dealer is honest, all correct parties complete the sharing phase.

Shamir Secret Sharing implements a $(k, n)$ threshold scheme with information-theoretic privacy [65]. To share a secret $s \in \mathbb{Z}_p$, the dealer samples a random degree-$(k-1)$ polynomial $f(X) \in \mathbb{Z}_p[X]$ with $f(0) = s$ and issues share $(i, f(i))$ to party $i$. Any set of $k$ shares can reconstruct the secret via polynomial interpolation, while any set of fewer than $k$ shares provides perfect indistinguishability.

We denote a shared value $s$ with threshold $k$ by $[s]_k$. Shamir sharing is homomorphic: from local operations on shares, one can derive shares of arithmetic combinations of secrets:

$$[a]_k + [b]_v = [a+b]_{\max(k,v)}, \qquad [a]_k \cdot \alpha = [\alpha a]_k, \quad (1)$$

These properties enable evaluation of arithmetic circuits on secret inputs [10], which we use to compute weighted vote tallies while hiding individual votes. The Secrecy property allows parties to secret share their vote without unwanted disclosure of information while the Termination property is required for the liveness of Privocracy's voting algorithm.

**Verifiable Secret Sharing** (VSS) schemes [5, 29, 39, 60] extends SS with publicly verifiable commitments that allow parties to check the consistency of their shares. These commitments prevent a malicious dealer from distributing inconsistent shares, and byzantine parties from providing incorrect values during reconstruction.

**Definition 3.2.** A VSS scheme satisfies SS and additionally:

- **Validity**: If the dealer is honest, the reconstructed value equals the shared secret.

A common instantiation adds Pedersen commitments [60]. Let $G$ be a cyclic group of prime order $q$ with generators $g, h$. To commit to $x \in \mathbb{Z}_q$, the dealer samples the blinding value $r \leftarrow \mathbb{Z}_q$ and publishes:

$$Commit(x) \equiv g^x h^r \mod q \quad (2)$$

Pedersen commitments are information-theoretically hiding, computationally binding, and additively homomorphic:

$$Commit(x) \cdot Commit(y) \equiv Commit(x + y) \mod q \quad (3)$$

Commitments can be extended to polynomial coefficients, enabling parties to verify that a received share is consistent with the committed polynomial and thus with the underlying secret.

**Asynchronous Verifiable Secret Sharing** (AVSS) [7, 21, 24] extend VSS to fully asynchronous networks, ensuring that shares are reliably delivered and remain consistent even if the dealer is faulty or selectively withholds messages.

**Definition 3.3.** An AVSS protocol satisfies VSS and additionally:

- **Agreement**: If any correct process completes the sharing phase with a share of a secret $s$, then no correct process completes with a share corresponding to a secret $s' \neq s$.
- **Totality**: If some correct process completes the sharing phase, then all correct processes eventually complete it.

Cachin et al. [21] an efficient AVSS construction in the computational model, tolerating up to $f < n/3$ Byzantine faults. This protocol uses Pedersen commitments [60] to bind shares and achieve $O(1)$ asynchronous rounds with $O(n^2)$ message complexity, establishing the practicality of AVSS for large-scale distributed systems.

AVSS allows Privocracy processes to coordinate the vote tally computation on an asynchronous network where processes can crash and exchange erroneous information. The Agreement property ensures that all correct processes will have shares corresponding to the same polynomial, while Totality will ensure Privocracy's liveness.

**Zero-Knowledge Proofs** (ZKPs) allow a prover to demonstrate the validity of a statement without revealing any additional information [43]. A common class of ZKPs are *Sigma protocols*, which are 3-move interactive proofs consisting of a commitment, a challenge, and a response. The verifier checks the resulting transcript to confirm correctness with soundness error negligible in $\lambda$.

Sigma protocols can be made non-interactive via the Fiat–Shamir heuristic [40], which derives the verifier's challenge by hashing the statement and the prover's commitment. They also support *OR-proofs* [32], enabling a prover to show knowledge of a witness to one of several statements without revealing which are true.

In our setting, these tools allow a voter to prove, in zero knowledge, that its committed vote is either 1 or 0, corresponding to an approval or disapproval, respectively. Such proofs prevent a malicious participant from submitting commitments to arbitrary values, thereby ensuring the integrity of the final tally while preserving ballot privacy.

**Broadcast** protocols allow a process to disseminate values while satisfying key correctness properties. Values are disseminated via *Broadcast* requests and received upon handling a *Deliver* event, which contains both the received value and its sender. **Byzantine Reliable Broadcast (BRB)** is a broadcast protocol that ensures all processes deliver the same values.

**Definition 3.4.** Byzantine Reliable Broadcast (BRB) satisfies the following properties [20]:

- **Validity:** If a correct process broadcasts $m$, every correct process eventually delivers $m$.
- **No duplication:** No message is delivered more than once.
- **Integrity:** If some correct process delivers $m$ with a correct sender $p$, then $m$ was previously broadcast by $p$.
- **Consistency:** No two correct processes deliver different messages.
- **Totality:** If some message is delivered by any correct process, every correct process eventually delivers a message.

BRB tolerates a maximum of $f < \frac{n}{3}$ faults [18] and can be implemented with a message communication complexity of $O(n^2)$ and a round complexity of $O(1)$ [2].

In our setting, although AVSS already ensures reliable delivery of secret shares, Privocracy additionally employs BRB to disseminate the dealer's integrity proofs consistently to all processes.

**Agreement** protocols allow processes to propose values and eventually reach a common decision. A process invokes a *Propose* request with a value; the primitive eventually triggers a *Decide* event with the agreed-upon result. Asynchronous Binary Agreement is a randomized and leaderless primitive where processes propose and eventually decide a binary value.

**Definition 3.5.** Asynchronous Binary Agreement (ABA) is an agreement protocol that satisfies the following properties:

- **Agreement:** No two processes decide differently.
- **Termination:** If all correct processes proposes, every correct process eventually decides.

- **Validity:** If every correct process proposes $v$, no correct process decides $\neg v$.

ABA tolerates up to $f < \frac{n}{3}$ faults [11]. Using a strong distributed coin to generate randomness [22], ABA achieves a communication complexity of $O(n^2)$ and round complexity $O(1)$ [1, 57].

ABA is commonly used as a building block for asynchronous Multivalued Consensus and Agreement on Common Subset, where proposals are broadcast and ABA instances decide which values contribute to the consensus [11]. In Privocracy, ABA plays a similar role, but instead of selecting proposals for a consensus, instances decide which sharings' secrets contribute to the final vote tally.

## 4 Access Policy

The purpose of an access policy is to ensure the correct and authenticated access to resources and information by the correct parties, which may be employees or collaborators of an organization, third-party users, etc.

Unlike well established access control mechanisms [63, 64], Privocracy can be used in policies that tolerate a modicum of human mistakes. Some users may assess harmful actions as innocuous, but as long as the majority of parties in the system are honest and competent, Privocracy can ensure the policy's properties.

Data owners define general access to files following the Discretionary Access Control mechanism. Owners have full permission to access a file and are able, through Access Control Lists (ACLs), to define fine-grained permissions for other users and groups. Data owners can decide to place files in a dedicated Privocracy group, where access to the files requires using the Privocracy voting procedure for all users that do not have explicit authorization to access the file. An institution using Privocracy can heavily incentivise granting access through the Privocracy group rather than explicit permissions in order to avoid the single point of failures common in traditional access control mechanisms.

### 4.1 Usage

Although Privocracy minimizes the need for privileged operations, its initial configuration requires limited administrative intervention. The only step that demands an administrator for the Privocracy setup is the creation of a dedicated system account, referred to as the `privocracy-user`. This account serves as the execution context for all Privocracy-related processes and provides a controlled privilege boundary for operations that require elevated access.

Once the account is created, the `privocracy-user` launches the Privocracy daemon, using the command described in table 1. The daemon operates as the coordination service of the system: it receives operation requests, distributes them to participating nodes for voting, aggregates their responses, and, upon approval, executes the corresponding commands with the permissions associated with the `privocracy-user`. This design ensures that no individual participant can unilaterally perform sensitive actions, while maintaining the capability to execute authorized commands on behalf of the collective.

**Table 1.** Execution commands for the Privocracy.

| Component | Privileges | Command |
|---|---|---|
| Daemon | privocracy-user | privocracy-daemon |
| Frontend | usr | privocracy <cmd> |
| Voter | usr | privocracy-voter |

File owners integrate their resources into this model by granting the `privocracy-user` access to specific files or directories. This is achieved using ACLs, for example in Linux:

```
setfacl -m u:privocracy-user:rw file.txt
```

In the above example, the user `usr`, who owns *file.txt*, gave read and write permissions for this file to the `privocracy-user`, and thus to the Privocracy daemon process running. By delegating access in this way, file owners retain full control over their data while allowing Privocracy to perform authorized operations transparently and securely.

Once configured, unprivileged users issue high privilege commands through the Privocracy frontend. As depicted in figure 2, when a user submits a command, the frontend constructs makes operation request to the daemon containing the command and the user's identifier, which logs it. The daemon then initiates a distributed voting round among designated approvers. Each node evaluates the request and contributes a hidden vote. The daemon then tallies the votes and executes the operation if the voting threshold was met. The operation executes under the privileges of the `privocracy-user` and its result is returned to the requester via the frontend interface.

This execution flow gives file owners a fine-grained control over the file permissions and the flexibility of not needing a central authorization mediator, as happens in Mandatory Access Control mechanisms. We envision it being integrated into existing administrative workflows by transparently replacing traditional privilege-escalation mechanisms, such as `sudo`, in regular workflow of the system.

## 5 Privocracy

Privocracy is a distributed, privacy-preserving access-control system designed to enforce collective authorization over critical operations. Rather than relying on a single administrator or a fixed Access Control List, Privocracy requires that multiple authorized entities approve a request before it executes, as depicted in figure 2.

When invoked, the Privocracy frontend intercepts the requested operation and forwards it to the Privocracy daemon on the corresponding Operational space instance. From the user's perspective, the interaction remains a simple command-line invocation; however, this action triggers the complete distributed e-voting protocol.

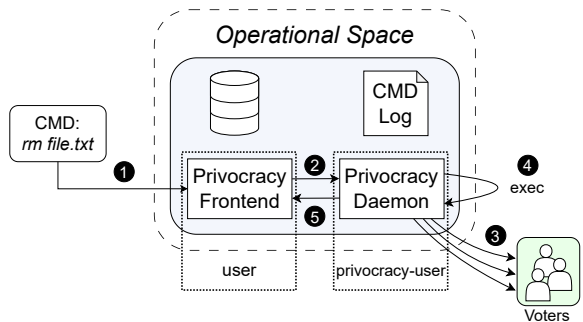

**Figure 2.** Operational flow of an access request through the Privocracy frontend and daemon. The frontend, which runs with the caller's permissions, directs commands to the daemon, which in turn runs with the elevated privileges of `privocracy-user`. The daemon triggers a vote to determine if the commands should be executed, logs requests and election metadata and, if approved, executes the commands. Finaly, the outputs are returned to the frontend.

Upon receiving the request, the Privocracy daemon initiates a voting procedure across the Voting Space. Voter instances are notified of the pending operation and prompt their respective human administrators to approve or reject the request. Each user casts a private binary vote (yes or no) via the Voter interface, which is then secret shared as 1 or 0, respectively, and cryptographically verified before aggregation. Once the voting round completes, the Privocracy daemon tallies the weighted results locally. If the cumulative vote weight exceeds the configured threshold Privocracy executes the command with the necessary privileges on the target Operational space instance. Regardless of the outcome, the Privocracy daemon logs the complete event into the Command Log. This information is then propagated across all Operational machines to ensure that every process converges to the same auditable state.

### 5.1 Voting

Privocracy voting procedure provides the following properties:

- **Integrity**: The final vote tally equals the weighted sum of the votes decided.
- **Synchronous Validity**: Upon GST, the final vote tally contains the votes of all correct processes.
- **Asynchronous Validity**: The final vote tally contains the votes of at least $n - 2f$ correct processes.
- **Secrecy**: If a correct process votes the adversary learns nothing about the vote value, bar what it can learn from the aggregate vote tally.

**Algorithm 1** Voting - Privocracy Daemon

```
 1: let V be the set of voters in the system;
 2:
 3: Uses:
 4:     Voter as vs [ ]
 5:
 6: Upon Init:
 7:     id ←ᴿ {0, 1}*
 8:     tallyShares ← ∅
 9:     threshold ← ⊥
10:     decided ← false
11:     weights ← ⊥
12:
13: Upon IssueCMD(issuer, command):
14:     threshold, weights ← loadConfig(issuer)
15:     trigger Log(id, command, issuer, self)
16:     for vᵢ in V: vs[vᵢ].Propose(id, issuer, command, weights)
17:
18: Upon vs[vᵢ].Deliver(partialTally, blindingTally, commit, weightSum):
19:     trigger Log(id, vᵢ, partialTally, commit)
20:     if IsValidVectorCommitment(partialTally, blindingTally, commit):
21:         tallyShares[< commit, weightSum >][vᵢ] ← partialTally
22:         trigger send(vᵢ, ACK)
23:         if #tallyShares[< commit, weightSum >] ≥ f + 1 and ¬decided:
24:             tally ← Interpolate(tallyShares[< commit, weightSum >])
25:             trigger Decide(tally ≥ threshold · weightSum)
26:             decided ← true
```

- **Termination**: If $n - f$ correct processes vote, the final vote tally is eventually computed.

Voters will issue manual intervention to vote $v \in \{0, 1\}$ to decide on whether a given command should be executed. Every vote $v_i$ has an associated weight $w_i$, such that the final vote tally is the weighted sum of all votes. If this sum exceeds a configurable threshold $t$, the command is executed.

Upon selecting $v_i$, its voter will secret share the value with a reconstruction threshold of $f + 1$. Because the network is asynchronous, voters may be byzantine, and manual intervention can be slow, the system must not wait for all votes. Thus, the voters will reach an agreement on which votes are considered for the final tally. Let $a_i$ be 1 if the vote is in the set of votes in the final tally and 0 otherwise, the condition to execute the command is:

$$\sum w_i v_i a_i \geq t \sum w_i a_i \qquad (4)$$

Every voter then computes a tally with their vote shares, rather than the votes themselves. The homomorphic properties of Secret Sharing guarantee that the reconstruction of the shares of the tally is equal to performing a tally on recovered votes.

$$\sum w_i [v_i]_{f+1} a_i = \left[\sum w_i v_i a_i\right]_{f+1} \qquad (5)$$

By recovering the tally shares rather than the votes themselves, the Privocracy daemon can accurately compute the tally without learning the individual votes of the voters.

Algorithm 1 represents the parts of the voting mechanism implemented by the Privocracy daemon. Upon receiving a

**Algorithm 2** IsBinary

1: let $g$ and $h$ be generators of group $G$, and $x$ and $r$ elements of the field $\mathbb{Z}_q$, where $q$ is a large prime.
2: let $C = g^x h^r \bmod q$ be the public Pedersen commitment to $x$
3:
4: **Objective**: Prove $x \in \{0, 1\}$
5:
6: **Upon** Prove$(x, C)$:
7:    $a \leftarrow x, \quad \bar{a} \leftarrow 1 - x$
8:    $r_a \xleftarrow{R} \mathbb{Z}_q \quad t_a \leftarrow h^{r_a}$
9:    $c_{\bar{a}} \xleftarrow{R} \mathbb{Z}_q \quad z_{\bar{a}} \xleftarrow{R} \mathbb{Z}_q$
10:    $t_{\bar{a}} \leftarrow h^{z_{\bar{a}}} C^{-c_{\bar{a}}} g^{c_{\bar{a}} \bar{a}}$
11:    $c \leftarrow Hash(C, t_a, t_{\bar{a}}) \bmod q$
12:    $c_a \leftarrow c \oplus c_{\bar{a}}$
13:    $z_a \leftarrow r_a + c_a r \bmod q$
14:    **output** $< t_a, t_{\bar{a}}, c_a, c_{\bar{a}}, z_a, z_{\bar{a}} >$
15:
16: **Upon** Verify$(C, < t_a, t_{\bar{a}}, c_a, c_{\bar{a}}, z_a, z_{\bar{a}} >)$:
17:    $c' \leftarrow Hash(C, t_a, t_{\bar{a}}) \bmod q$
18:    **output** $c' = c_a \oplus c_{\bar{a}} \wedge (h^{z_a} \equiv t_a C^{c_a} g^{-c_a a} \bmod q) \wedge (h^{z_{\bar{a}}} \equiv t_{\bar{a}} C^{c_{\bar{a}}} g^{-c_{\bar{a}} \bar{a}} \bmod q)$

**Algorithm 3** Voter

1: let $f(\delta)$ be the upper bound on the time difference between votes by correct processes;
2: let $V$ be the set of processes in the system;
3:
4: **Uses:**
5:    **Asynchronous Verifiable Secret Sharing** as avsss[ ]
6:    **Asynchronous Binary Agreement** as abas[ ]
7:    **Byzantine Reliable Broadcast** as brbs[ ]
8:
9: **Upon** Init:
10:    $voteShares \leftarrow \{(v_i, \bot) | v_i \in V\}, \quad blindingShares \leftarrow \{(v_i, \bot) | v_i \in V\}$
11:    $decs \leftarrow \{(v_i, \bot) | v_i \in V\}, \quad vecCommits \leftarrow \{(v_i, \bot) | v_i \in V\}$
12:    $weights \leftarrow \emptyset, accepting \leftarrow true$
13:
14: **Upon** Propose$(cmdId, v_i, cmd, ws)$:
15:    $weights \leftarrow ws$
16:    $vote \leftarrow$ ManualVote$(v_i, cmd, ws)$
17:    $blinding \xleftarrow{R} \mathbb{Z}_q$
18:    $commit \leftarrow g^{vote} h^{blinding} \bmod q$
19:    $proof \leftarrow$ IsBinary.Prove$(vote, commit)$
20:    **trigger** avsss$[self]$.Share$(vote, blinding)$
21:    **trigger** brbs$[self]$.Broadcast$(proof)$
22:
23: **Upon** avsss$[v_i]$.Deliver$(vShare, bShare, vCommit)$ and brbs$[v_i]$.Deliver$(proof)$:
24:    **trigger** Log$(cmdId, v_i, vShare, bShare, vCommit)$
25:    $voteShares[v_i] \leftarrow vShare$
26:    $blindingShares[v_i] \leftarrow bShare$
27:    $vecCommits[v_i] \leftarrow vCommit$
28:    **if** $accepting$ and IsBinary.Verify$(vCommit[0], proof)$:
29:       **trigger** abas$[v_i]$.Propose$(true)$
30:
31: **Upon** abas$[v_i]$.Decide$(b)$:
32:    $decs[v_i] \leftarrow b$
33:
34: **Upon** $\#\{(\cdot, true) \in decs\} \geq n - f$ and $elapsed > f(\delta)$:
35:    $accepting \leftarrow false$
36:    **for** $v_i \in \{(v | (v, \bot) \in voteShares\}$ **do** abas$[v_i]$.Propose$(false)$
37:
38: **Upon** $\#decs = n$ and $\{v | (v, true) \in decs\} \subseteq \{v | (v, \neg \bot) \in voteShares\}$:
39:    $partialTally \leftarrow \sum_{v_i \in V} (voteShares[v_i] \cdot weights[v_i] \cdot decs[v_i])$
40:    $blindingTally \leftarrow \sum_{v_i \in V} (blindingShares[v_i] \cdot weights[v_i] \cdot decs[v_i])$
41:    $commit \leftarrow \sum_{v_i \in V} (vecCommits[v_i])^{weights[v_i] \cdot decs[v_i]}$
42:    $weightSum \leftarrow \sum_{v_i \in V} weights[v_i] \cdot decs[v_i]$
43:    **trigger** Deliver$(partialTally, blindingTally, commit, weightSum)$

command, the Privocracy daemon loads its configurations, based on the issuer and the specific permissions of the Operational space instances, and logs the operation in the synchronized log. The requested command, along with the issuer and voter weight, are then sent to the voters so they run an election on whether the operation should be executed.

Eventually, the voters will respond with shares of the final tally, from whose reconstruction the Privocracy daemon can assert if the command can be run. Some voters can be byzantine and can therefore send incorrect values to the Privocracy daemon. Naively using these incorrect values can lead to an incorrect interpolation of the vote tally. Instead, voters also send shares of an auxiliary *blinding* polynomial and a Pedersen [60] vector commitment of the joint coefficients of the final tally and *blinding* polynomials. Because any two shares associated with matching commitments correspond to the same underlying polynomial, and all honest voters disseminate identical commitments with valid shares, the daemon can recover the tally by interpolating any f+1 shares that share the same commitment.

One challenge in the algorithm is determining if a malicious voter is sharing a valid vote. If a voter unduly shares a $v \notin \{0, 1\}$, it will have a disproportional representation in the final vote tally. The *IsBinary* functions, presented in algorithm 2, generate and verify proofs that the votes are either 1 or 0 based on Pedersen commitments [60] used in the AVSS [21] protocol. This predicate is a Sigma protocol with an Or composition [32] and made non-interactive through the Fiat-Shamir heuristic [40].

Algorithm 3 represents the parts of the voting algorithm implemented by the voters. Each Voter receives the command to be executed, along with who issued the command and the weights used in the vote. From this information, the Voter requests manual intervention to vote on whether the command should succeed. Afterwards, the protocol is similar to Ben-Or, Kelmer, and Rabin's [11] implementation of Agreement on Common Subset [8]; but instead of using a Byzantine Reliable Broadcast to disseminate their values, the voters will use Asynchronous Verifiable Secret Sharing (AVSS) instances to reliably disseminate their secret shares. Furthermore, instances will also wait for a timeout (as $f(\delta)$) representing the voting period. The protocol proceeds in the following steps:

1. An Asynchronous Binary Agreement (ABA) instance is created for each process.
2. Each process requests a manual vote $v$ and generates a proof for $v \in \{0, 1\}$.
3. Each process AVSS-Shares their vote and BRB-Broadcasts their proof.

4. When a correct process AVSS-Delivers a share and BRB-Delivers the corresponding proof. If the proof is valid, the process proposes *true* (accept) in the sender's ABA instance.
5. When a correct process decides $n - f$ values *true* from the ABA instances and the voting timeout is exceeded, it proposes *false* (reject) in all ABA instances where it did not yet propose;
6. When all ABA instances finish, processes compute the weighted tallies of with their shares according to equation 5, and send the results to the daemon that issued the command. Additionally, processes also send the weighted blinding polynomial shares, the vector commitment needed for the daemon to verify the validity of the tally shares, and the sum of the weights of the processes whose votes were accounted for.

**Argument for Integrity:** Upon concluding the ABA rounds all correct processes will perform the weighted sum of the decided shares. These correspond to the same secrets across voters, as ensured by the *Agreement* property of AVSS. Regardless of whether their votes were decided or not by the ABA instances, the correct processes will compute the weighted vote tallies with the decided share values and send these results to the daemon; ensuring the majority of tally shares comes from correct processes and thus allowing a correct reconstruction of the result.

**Argument for Synchronous Validity:** If the network is behaving synchronously and all correct voters perform the manual vote in due time, all vote sharings will be received by correct processes before the voting timeout is triggered, thus ensuring all correct processes will accept each other's votes in the corresponding ABA instances. As per the *Validity* property of ABA, if all correct processes agree on each other's votes, the adversary is unable to lead the ABA instances to decide otherwise.

**Argument for Asynchronous Validity:** In the ABA instances, correct processes only vote negatively after having at least $n - f$ positive results from other instances. From these positive result, at most $f$ were proposed by Byzantine processes; hence, the decided values contain at least $n - 2f$ correct proposals.

**Argument for Secrecy:** The voting process uses Shamir Secret Sharing as the underlying protocol, ensuring no information is gathered from any number of shares bellow the reconstruction threshold. This reconstruction threshold is greater than the adversary threshold, ensuring they are unable to learn any information from the shares observed. The AVSS protocol and algorithm 2 use Pedersen commitments to ensure the integrity of the shares. These commitments ensure blinding under information-theoretic security. Therefore, the adversary is unable to learn anything about individual votes from corruptions to the Voter instances, regardless of their computational power.

During reconstruction, the Privocracy daemon receives the partial vote tallies. Because every share is a random sample from a uniform distribution from $\mathbb{Z}_q$, the sum of shares from the votes is itself uniform in $\mathbb{Z}_q$, such that the adversary is unable to learn anything from the partial tallies reconstruction, as long as more than one correct process has their vote decided, which in turn is ensured by the *Asynchronous Validity* property.

**Argument for Termination:** The ABA instances are a barrier in the procedure, given that all must terminate before the voting can end. Furthermore, their *Termination* property requires all correct processes to propose a value.

If a correct process votes, the sharing of its vote will terminate as per the *Termination* property of AVSS. Every sharing terminated will lead all correct processes to accept the corresponding vote in an ABA instance, unless they have proposed a rejection before; triggering the *Termination* condition for the ABA instances. Therefore, even if only $n - f$ correct processes vote while all Byzantine voters abstain, the correct processes will finish enough sharings in order to accept $n - f$ ABA instances. As soon as these ABA instances have agreed to accept the proposed votes, correct processes can reject the remaining ABA instances, triggering their *Termination* condition, thus overcoming the ABA instances barrier.

**Corollaries:** From the properties of the voting system, we can infer the following statements regarding the adversary's ability to decide the end result of an election.

**Corollary 5.1.** *Upon GST, if the adversary has less than half of the cumulative vote weight, it is unable to obtain a majority vote if all correct processes vote against the adversary's action.*

It follows from the *Synchronous Validity* property that upon GST all votes cast will contribute to the final vote tally. If all correct processes vote on a value, it will be the majority vote regardless of adversarial intervention, as long as the adversary has less than half to total vote weight. In practice, we expect this scenario to be the norm, given the potentially prolonged timeouts the system can use.

**Corollary 5.2.** *If the cumulative vote weight of the adversary is less than every subset of $n - 2f$ correct processes, then the adversary is unable to obtain a majority vote if all correct processes vote against its action.*

It follows from *Asynchronous Validity* that $n - 2f$ correct processes' votes are tallied. If the votes from any such subset of correct processes have greater cumulative weight than the adversary's, it cannot unilaterally decide the end result. From this we can infer a further statement:

**Corollary 5.3.** *If all voters have equal weights, the adversary is unable to unilaterally decide the output of a voting procedure.*

If vote weights are equal across all processes, the adversary is unable to achieve a majority vote, given that any vote tally will have the votes of at least $f + 1$ correct processes. Even if the weights are not uniform, it is expected (although not guaranteed by the system model) that in a deployed system the byzantine corruptions will be more prominent in lower weighted processes if the weights are attributed based on the responsibility and role of the associated human party.

In sum, even though there is no guarantee the adversary is unable to decide an election, to do so it must either have the vote majority in the system, or have a disproportionate vote weight while correct processes fail to vote within the allotted time.

**Complexity Analysis:** For $n$ processes, the message complexity of the algorithm is dominated by BRB and AVSS, both of which boosting a message complexity of $O(n^2)$. Given that for each vote, requires $n$ instances for each voting sequence, the overall message complexity of algorithm is $O(n^3)$.

If the network is not overloaded, the latency of the protocol is dominated by the number of sequential communication rounds. BRB, AVSS, and ABA have an expected constant number of communication rounds and therefore a latency $O(1)$. However, running $n$ ABA instances and expecting all to finish induces a $O(\log n)$ latency [9, 37], hence this is the complexity of the voting algorithm.

## 5.2 Audits

If a command issued through Privocracy damages the system or unduly discloses information, its responsible should not only be the command issuer, but also the parties that democratically allowed the command's execution. Similarly, if the system is hindered through inaction, the parties that prevented a solution from being enacted should be held responsible. To achieve this, we implement a secure auditing mechanism, such that the votes of a command executed within a given timeframe can be revealed. During the voting procedure (algorithm 3) voters locally log the shares output by the AVSS sharing. These shares are secret and are not revealed except during an audit. To perform an audit, a user performs the simple command:

```
privocracy –audit <op_id>
```

The operation identifier received as argument references the operation to be audited and can be retrieved from the command logs in the Operational space instances.

Issuing an audit requires a majority vote. This process differs slightly from regular voting (§ 5.1) given that all voters have equal weights, ensuring the adversary is unable to force a vote disclosure or unduly prevent an audit, as follows from corollary 5.3.

During an audit voters divulge the local shares associated with op_id, along with their commitments, from which the daemon can reconstruct each individual share. Even if the byzantine voters refuse to divulge their logged shares, the correct processes' shares are enough to reconstruct the votes. AVSS ensures that all correct processes have equal commits for the shares of a given ballot. Upon receiving $f + 1$ shares with the same commitment, the daemon is able to reconstruct the vote value knowing all shares correspond to the same polynomial.

## 5.3 Emergency Votes

In real world deployments, specific operational contexts demand faster decision-making than the standard collective authorization workflow allows. Examples include emergency maintenance and critical infrastructure recovery. To address these scenarios, Privocracy introduces *Emergency Votes*, a latency-optimized variant of the normal voting procedure that maintains privacy and verifiability guarantees while allowing early termination of the voting round under well-defined conditions.

Emergency votes use the same underlying cryptographic and privacy-preserving mechanisms as standard votes. The key distinction lies in the introduction of a *dual-threshold decision rule*, which enables the system to decide whether to receive all vote shares.

Let $t \in {]}0, 1]$ denote the approval threshold (i.e., the fraction of the total weighted votes required to authorize an operation). Under normal conditions, Privocracy waits for all eligible votes before computing the final tally, if a timeout was not triggered. In an emergency vote, however, Privocracy evaluates the partial tally once it has received votes from at least $t_e$ of the collective. Where, $t_e$ corresponds to:

$$t_e = 1 - \frac{1-t}{2} \quad (6)$$

This defines an *early evaluation frontier*, the point at which the system has observed enough partial information to make a safe decision in most cases.

At $t_e$, Privocracy distinguishes three cases:

1. The current partial tally already exceeds the threshold $t$, implying that the operation will necessarily be approved even if all remaining votes are negative. Privocracy immediately authorizes execution.
2. Even if all remaining votes are positive, the operation cannot reach $t$. Privocracy promptly rejects the request.
3. The outcome remains undecidable with the current information. Privocracy continues to await the remaining votes until the standard threshold condition is resolved.

This mechanism substantially reduces decision latency in time-critical cases, as it prevents the system from stalling on unresponsive or delayed voters once the outcome is logically determined. Because early stopping relies on partial information, it is restricted to explicitly triggered emergency scenarios rather than normal operation - invoked via:

```
privocracy -emergency <command>
```

This allows command issuers to intentionally trade a controlled reduction in voter representation for faster decisions, while still preserving overall system security and maintaining operational continuity.

### 5.4 Vote Delegation

In standard voting rounds, the last few votes can significantly slow down the decision-making process, potentially delaying the execution of critical operations, as seen in the aforementioned subsection. To mitigate this latency, Privocracy introduces *Vote Delegation*, a mechanism that allows voters to rely on the previous knowledge of a trusted social circle to cast a fallback vote on their behalf [16].

Each voter can establish a trust relationship with other voters, represented as a directed *trust graph* $G = (V, E)$, where $V$ is the set of voters and $E$ encodes trust edges. Edges can carry weights reflecting confidence, and transitive delegation is possible: for example, if $A \to B$ and $B \to C$, $A$ may delegate their vote through $B$ to $C$ if necessary. Delegation is triggered automatically when a voter fails to cast a vote within a timeout period.

Formally, we define a delegation mapping $D : V \to \mathcal{P}(V)$ that maps each voter to a set of delegate voters. When a timeout occurs, the Privocracy daemon uses $D$ to determine the fallback vote for the absent voter by choosing one of the possible relations.

All trust graphs are shared across Operational space instances and can be updated over time, especially after an audit, where a voter can find if its trusted voter has acted maliciously, so that it can remove or adjust the corresponding trust edges. This ensures that delegation remains adaptive and reflective of historical reliability.

Vote delegation relaxes strict voter accountability because a delegate's ballot may diverge from the delegator's original intention. At the same time, it improves liveness: delegations allow decisions to proceed even when some voters are slow or unavailable, as the delegate casts a single vote weighted by all incoming delegations [17].

While vote delegation can increase decision accuracy under suitable conditions [46], excessive reliance on delegation can undermine democratic balance. Empirical studies of real-world liquid democracy platforms show that delegations tend to concentrate voting power and create super-voters [49]. Theoretical analyses further demonstrate that local or naive delegation mechanisms may perform significantly worse than direct voting and may even amplify errors [23, 25]. Additional structural issues, like including delegation cycles, collective abstentions, and logically inconsistent collective outcomes, have also been identified in foundational work on delegable proxy voting [30].

To mitigate these effects, our voting procedure (algorithm 3) enforces a minimum threshold of direct participation (through the proposals required to terminate the ABA instances). Additionally, Privocracy allows setting an upper bound on each voter's cumulative delegated weight. A delegate only receives an incoming delegation if doing so keeps its total weight below a configurable percentage of the voters aggregate weight, thereby limiting the emergence of super-voters [46]. These mechanisms maintain timely collective authorization while reducing vulnerability to power concentration, strategic delegation, and unresponsive participants inherent to vote delegation.

By combining social trust, transitive delegation, and timeout-based triggers, Privocracy ensures that collective authorization remains both practical and resilient, even when some participants are slow or unavailable.

### 5.5 Privacy and Leakage Analysis

Privocracy is designed so that its sensitive operational component, the daemon, reconstructs the tally, and evaluates whether the approval threshold is met - executes inside a Trusted Execution Environment (TEE). When correctly deployed within a TEE, the daemon reveals only a single public bit indicating whether the proposal is approved. It does not expose the tally value $T$, the participation subset $S \subseteq \{1, \ldots, n\}$, or any intermediate computation state. Under this intended deployment model, the adversary learns nothing beyond the final decision of an unknown subset of voters, and Privocracy provides zero vote leakage.

The remainder of this subsection analyzes what an honest-but-curious adversary *could* learn in a weaker deployment where the TEE is absent or not properly enforced. In such settings the adversary reconstructs the final tally $T$, and may try to infer the votes:

$$T = \sum w_i v_i a_i \qquad (7)$$

where $v_i \in \{0, 1\}$ is the private vote of user $i$, $w_i$ is that user's public weight, and $a_i$ indicates whether user $i$ was selected in that round. The adversary does not know the exact subset of voters $S$, but only that its size lies in $|S| \in [n - f, n]$. This model enables the vote-leakage phenomena over multiple rounds issued by the same user over the same operational command, as described below and should be interpreted as a worst-case adversarial scenario for non-TEE deployments.

**Information Leakage per Round:** Let $V = (v_1, \ldots, v_n)$ denote the vector of votes and let $T_r$ be the tally of round $r$. Each tally restricts the set of compatible vote vectors. Writing $\mathcal{T}_r$ for the set of possible tallies given the allowed subset sizes, a classical information-theoretic upper bound on the leakage in round $r$ is given by:

$$L_r = I(V; T_r) \le \log_2 |\mathcal{T}_r| \qquad (8)$$

Under the assumption of independent uniform votes, $H(V) = n$ bits, so leakage over all rounds cannot exceed $n$ bits.

**Monotonicity:** Leakage is monotone since each tally adds a new linear constraint and can only shrink the space of vote vectors consistent with prior observations.

$$L_r \geq L_{r-1} \quad (9)$$

**Saturation:** There exists a finite $R_{\max} \leq n$ such that for all $R \geq R_{\max}$.

$$L_R = L_{R_{\max}} \quad (10)$$

Beyond this point, additional tallies do not further reduce uncertainty: the adversary has learned all information extractable from the weight structure and sampling process.

**Effect of Weights:** The weight distribution determines the severity of leakage:

- **Uniform weights ($w_i = w$):** Tallies reveal only the number of approving votes in the participating subset. Because all voters have identical weights, they are indistinguishable under the tally, yielding $k$-anonymity [67] with $k = n$. Leakage is therefore minimal, unless all voters cast the same votes, in which case no entropy exists.
- **All weights distinct:** Each tally is a unique linear combination of weighted votes. Over multiple rounds, these constraints can fully determine $V$, and leakage approaches full disclosure.
- **Grouped weights:** If weights form $m$ equivalence classes, users within each class are indistinguishable. Leakage occurs at the group level, providing $k$-anonymity within each weight group unless all group members vote identically.

In summary, the leakage analysis above applies only to non-TEE deployments. In the intended design, where the tally-reconstruction logic is isolated inside a TEE and only a single approval bit is released, none of these leakage channels are available, and vote privacy is preserved.

## 6 Implementation

All protocols were implemented in Go, leveraging the `Circl` library for group operations and discrete-log equivalence proofs [27]. All group operations are performed over the `ristretto255` elliptic curve [35]. The underlying Byzantine Reliable Broadcast (BRB), Asynchronous Binary Agreement (ABA), and Asynchronous Verifiable Secret Sharing (AVSS) implementations follow the algorithms described in [2], [1], and [21]. The $2f$-unpredictable strong coin used in the ABA algorithm follows the construction presented in [22]. Privocracy daemons communicate directly with the voters via a REST API and synchronize the operation log across machines using Apache Kafka [51], ensuring all Privocracy daemons converge to a consistent, auditable state, and providing fault-tolerant coordination without relying on synchronous communication.

**Table 2.** Testbed specifications

| Component | Specification |
| --- | --- |
| CPU | 2× AMD EPYC 9124 16-Core Processor (64 cores total with hyperthreading, 3.7 GHz max) |
| Voter Processing | 8 cores per Voter |
| RAM | 265 GB |
| NUMA Interconnect | AMD Infinity Fabric |
| Network | 2× 25 GbE (bonded, 802.3ad LACP) |

For manual vote input, the system displays native popup dialogs across operating systems, prompting administrators to approve or reject pending operations.

## 7 Evaluation

To assess the practical performance of Privocracy, we measured the latency of executing operations under varying threshold sizes and network conditions. These experiments simulate the impact of increasing the number of voters, which scales with the threshold using the standard relation $n = 3f + 1$. As the number of participants grows, both communication overhead and vote aggregation complexity are expected to influence overall operation latency. Our experiments will show that for reasonable number of voters the end-to-end latency of Privocracy is not prohibitive for practical deployments and that users of the system can operate it using commodity hardware.

All experiments were executed on a Docker Swarm cluster composed of identical machines. The configurations of each machine are summarized in table 2. Network latency between voters was artificially introduced using the Linux Traffic Control (`tc`) utility with the `netem` qdisc to emulate representative deployment scenarios. Voters are assumed to respond instantaneously, allowing the benchmark to isolate only the coordination and aggregation overhead of the system, and decided to comprise three representative network conditions: 5 ms (low), 40 ms (medium), and 150 ms (high) of latency between voters. Moreover, the threshold parameter $f$, presented in the following sections, is not used to simulate Byzantine behavior in these experiments; rather, it serves solely as a mechanism for increasing the number of voters.

### 7.1 End-to-End Latency

In figure 3 we present the mean operation latency as a function of the threshold $f$, along with the 25th–75th percentile (lighter colour) and the minimum/maximum durations (lightest colour). The linear-scale plot highlights the absolute values and the growing trend on time duration, while the logarithmic-scale plot emphasizes the convergence trend.

As the number of voters increases, mean operation latency rises across all network conditions due to the additional messages, coordination effort, and aggregation work required. For small thresholds, latency is dominated by network delay;

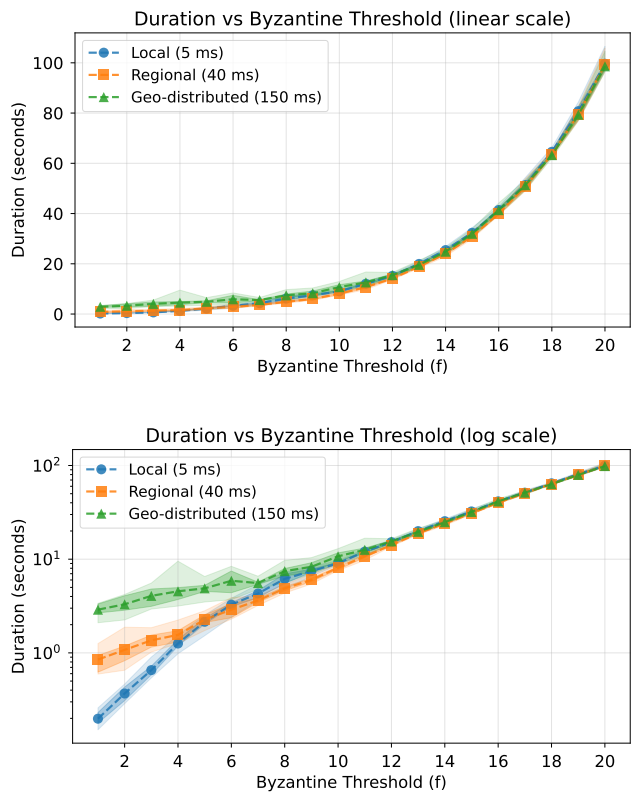

**Figure 3.** Latency statistics for operations under varying thresholds, with the number of voters set to $3f + 1$ (from 4 to 61). The top figure shows results on a linear scale, while the bottom uses a logarithmic scale to highlight the convergence of durations across different network latencies as the threshold increases. As the number of Voters increases, the impact of the network latency between Voters in the end-to-end latency of an election decreases.

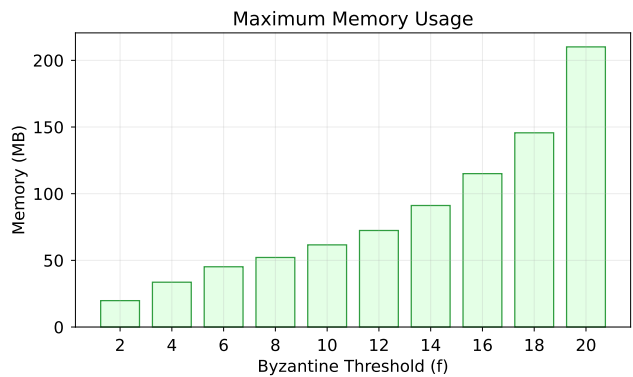

**Figure 4.** Maximum memory usage across Byzantine thresholds. As the number of voters increases, so does the memory requirements of the protocol, but not enough so running these systems is intractable for commodity hardware in elections with dozens of voters.

### 7.2 Memory Requirements

As expected, the maximum memory consumption also increases with the threshold, as depicted in figure 4. Until reaching 25 voters, memory usage remains below 50 MB, and even at the highest threshold tested it grows only to slightly above 500 MB. This increase reflects the additional state required for storing votes, partial tallies, and intermediate data structures as the number of voters scales. Memory usage is effectively independent of network latency, since all vote-related data is maintained locally. Overall, the memory footprint remains well within the capabilities of commodity hardware, indicating that Privocracy can scale to substantially larger configurations without imposing significant memory demands.

Even at the highest threshold tested, resource usage remains comfortably below hardware limits, indicating that the system can support substantially more voters without exhausting compute or memory resources.

### 7.3 Voting Patterns

Frequent elections over short periods of time, such as those resulting from the use of Privocracy, induce voter fatigue [50], which in turn reduces voter turnout. We analysed the Wikipedia adminship elections dataset [54, 55], which contains nearly 2 800 elections, more than 10 000 users (voters and votees), and over 198 000 votes. While the number of voters differs across the adminship elections, their timing patterns, depicted as a cumulative distribution function in figure 5, are consistent for both large and small elections and highlight a consistent structure across elections.

We observe that voting activity is highly front-loaded: participants who intend to vote tend to do so early, creating a rapid initial rise in the cumulative vote percentage. As seen in the median curve, elections surpass the 50% completion for example, at $f = 1$, mean durations range from approximately 0.2 s (5 ms) to 2.9 s (150 ms).

As $f$ increases, the curves for different network latencies converge, showing that coordination and computation overheads begin to dominate over network delay. This produces a superlinear growth pattern, highlighting the trade-off between scalability and performance: larger systems can tolerate more participants, but at the cost of higher latency per operation.

Overall, these results demonstrate that while system overhead increases with the number of voters, the dominant factor in operation latency is expected to be the time required for participants to cast their votes, rather than the coordination or aggregation performed by the system.

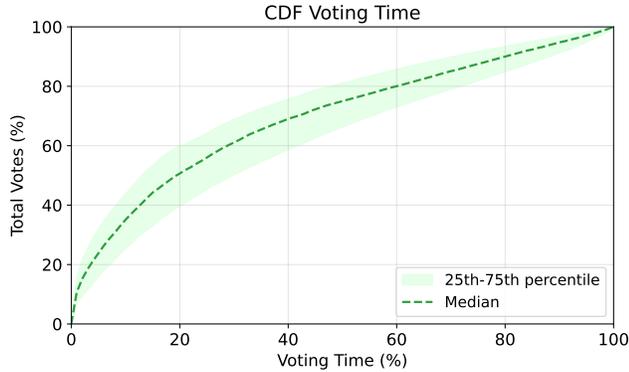

**Figure 5.** Cumulative vote progression over normalized election time for Wikipedia adminship elections, showing rapid early vote accumulation followed by a pronounced slowdown in later stages.

mark within roughly the first 20% of their duration, indicating that the majority of engaged voters act shortly after the election opens. This momentum continues initially, with the 75% threshold typically reached around the halfway point of the voting window. However, once this threshold is crossed, a clear deceleration in voting activity emerges. The tail end of the curve flattens considerably, showing a significant decrease in the voting rate during the latter half of the election and suggesting a diminishing pool of voters who either have not yet seen the election or are less motivated to participate. In Privocracy, this pattern suggests that most meaningful participation is likely to happen early, with prolonged voting periods offering limited additional benefit.

From this data we gather that having high voting timeouts has diminishing returns. Using the vote delegation (§ 5.4) feature, the system can be configured to wait reduced timeouts while still garnering most of the votes and still account for the few stragglers' votes through their delegation. Furthermore, this data shows that during emergency voting (§ 5.3) the system can quickly gather enough votes for a preliminary tally whose result can allow quick actions during emergencies.

## 8 Related Work

**Voting Systems:** Electronic voting has been extensively studied by the cryptography community [14, 31, 45], with emphasis on achieving both ballot privacy and end-to-end verifiability [12]. Privacy ensures that no party can learn how an individual voted, while verifiability enables voters and auditors to confirm that all ballots were correctly recorded and tallied.

A central design technique in many e-voting systems is to *unlink* ballots from voter identities, making it computationally infeasible to associate a cast ballot with its issuer [3, 44, 48, 68]. Such unlinkability is often achieved through primitives such as blind signatures [26] or mix-nets [28]. However, once unlinking is performed, later re-association between ballots and voters is impossible, limiting the system's ability to perform selective audits or accountability procedures. In contrast, our design preserves a controlled link between ballot shares and voters, enabling robust auditability while maintaining secrecy of vote contents.

Another line of work uses additively homomorphic encryption [59, 69] to support privacy-preserving tallying [34, 36]. Homomorphically encrypted ballots can be aggregated without revealing individual votes, and the final aggregate is decrypted to obtain the election result. These systems, however, require trust in the entities holding the decryption key material: if these parties decrypt individual ciphertexts rather than the aggregate, ballot privacy is compromised. This vulnerability persists even against honest-but-curious adversaries.

Secret-sharing and multiparty-computation (MPC) based solutions distribute trust among various instances, thus providing information-theoretic privacy and eliminating single points of failure [33, 42, 66]. While secret-sharing protocols incur relatively high communication complexity [21], their computational costs are typically low compared to homomorphic-encryption-based approaches, making them practical for elections with dozens of voters.

Most e-voting literature focuses on single-shot elections involving large electorates and assumes relatively synchronous networks and stable infrastructure. Our setting differs substantially: we require a voting system that operates reliably over long periods, supports frequent elections with modest numbers of participants, and remains robust under asynchrony and machine failures. Our solution therefore addresses a distinct systems-level challenge—designing a voting mechanism that is both cryptographically sound and dependable in a distributed, long-running deployment.

**Access Control:** Traditional UNIX systems rely on Discretionary Access Control (DAC), which prioritizes ease of use and user-level flexibility over strong security guarantees [63]. In DAC, an administrator with `sudo` privileges wields unrestricted authority, and processes automatically inherit the full permissions of the invoking user. As a consequence, the internal security of the system depends on the honesty and operational care of privileged administrators. This model is particularly vulnerable to Trojan Horse attacks [63]: a seemingly benign program executed by an administrator can exploit inherited privileges to read or modify sensitive system resources without further restriction. Although Privocracy does not eliminate this attack vector entirely, its requirement that commands be observable by a large set of Voters increases scrutiny and reduces the likelihood of a successful undetected compromise.

Mandatory Access Control (MAC) and Role-Based Access Control (RBAC) [64] were introduced to address these limitations of DAC. In both models, a process does not automatically inherit all of a user's privileges; instead, permissions are determined by system-wide policies or by roles explicitly activated by the user. These restrictions mitigate Trojan Horse attacks by preventing arbitrary programs from exploiting a user's full authority. However, defining and maintaining MAC or RBAC policies is challenging in large organizations with many users and heterogeneous resource requirements. Such policies may also unduly restrict user flexibility. For this reason, MAC and RBAC are typically deployed alongside DAC [63], and can be complemented by systems such as Privocracy that support more flexible distributions of trust.

An alternative approach to distributing trust is provided by two-person integrity mechanisms such as ISE-T [61], which require two distinct administrators to authorize privileged operations. This design protects against unilateral malicious actions but reduces liveness, since progress depends on the correctness and availability of both parties. Privocracy generalizes this idea to an arbitrary set of participants, ensuring safety and liveness as long as the proportion of adversarial actors remains below a configurable threshold.

COLBAC [41] proposes a decentralized access-control model in which authorization decisions are made democratically by all members of an organization. Like Privocracy, COLBAC aims to eliminate reliance on a single trusted administrator. However, whereas Privocracy extends Linux's DAC model with collectively authorized privilege delegation, COLBAC introduces additional abstractions, such as collective resources, immutable logs, and Emergency Tokens, that enable broader generality at the cost of greater conceptual and operational complexity. COLBAC handles emergencies by granting selected parties temporary unrestricted access via Emergency Tokens, while Privocracy retains a more horizontal permission structure by employing Emergency voting rounds.

Implementations of COLBAC [56] are closed-source and rely on Pluggable Authentication Modules (PAM) to intercept authorization checks for privileged commands such as sudo, replacing them with a collective approval protocol. This eliminates the need for traditional system administrators. Privocracy adopts a less intrusive approach: it allows the creation of multiple `privocracy-users` with fine-grained, collectively governed permissions while preserving compatibility with well-established access-control mechanisms [63]. Unlike COLBAC [41, 56], Privocracy incorporates a privacy-preserving electronic voting scheme, thereby safeguarding the anonymity of participating voters, and allows voters to have different, yet configurable, weights, thus improving the flexibility of the voting system.

## 9 Conclusion

This paper introduces Privocracy, an access control mechanism that extends the Linux default discretionary access control to distribute trust in the execution of commands that require access to privileged files, through the use of an e-voting procedure. In doing so, Privocracy reduces the single points of failure that would result from a high number of parties having unrestricted access to sensitive resources. Our open-source prototype implementation and its evaluation show the applicability of Privocracy for elections with dozens of voters, each of which running commodity hardware.

Privocracy opens a future research direction on the effects of regular voting on systems' users. Voting fatigue is a long-standing concern in elections [50]; a measurement and potential mitigation of its effects on an access control system with frequent voting is surely needed to better understand the trade-offs in distributing trust for human operators.

## Acknowledgements

This work was supported by Fundação para a Ciência e a Tecnologia, I.P. (FCT) through the grants NOVAID-B336-ParSec/BI/03 and NOVAID-B360-ParSec/BI/04 and through the PhD scholarship 2025.15439.PRT. The work was also supported by the NOVA Laboratory for Computer Science and Informatics (NOVA LINCS) research unit (UID/04516/2025) and by the Portuguese Recovery and Resilience Plan (PRR) under the measure RE-C05-i08.m04, through the "ParSec: Astronomically Improving Parliamentary Cybersecurity through Collective Authorization" project (2024.07643.IACDC).

## References

[1] Ittai Abraham, Naama Ben-David, and Sravya Yandamuri. 2022. Efficient and Adaptively Secure Asynchronous Binary Agreement via Binding Crusader Agreement. In *Proceedings of the 2022 ACM Symposium on Principles of Distributed Computing* (Salerno, Italy) *(PODC'22)*. Association for Computing Machinery, New York, NY, USA, 381–391. https://doi.org/10.1145/3519270.3538426

[2] Nicolas Alhaddad, Sourav Das, Sisi Duan, Ling Ren, Mayank Varia, Zhuolun Xiang, and Haibin Zhang. 2022. Balanced Byzantine Reliable Broadcast with Near-Optimal Communication and Improved Computation. In *Proceedings of the 2022 ACM Symposium on Principles of Distributed Computing* (Salerno, Italy) *(PODC'22)*. Association for Computing Machinery, New York, NY, USA, 399–417. https://doi.org/10.1145/3519270.3538475

[3] Myrto Arapinis, Véronique Cortier, Steve Kremer, and Mark Ryan. 2013. Practical Everlasting Privacy. In *Principles of Security and Trust*, David Basin and John C. Mitchell (Eds.). Springer Berlin Heidelberg, Berlin, Heidelberg, 21–40.

[4] Algirdas Avizienis, Jean-Claude Laprie, Brian Randell, and Carl Landwehr. 2004. Basic Concepts and Taxonomy of Dependable and Secure Computing. *IEEE Trans. Dependable Secur. Comput.* 1, 1 (Jan. 2004), 11–33. https://doi.org/10.1109/TDSC.2004.2

[5] Michael Backes, Aniket Kate, and Arpita Patra. 2011. Computational verifiable secret sharing revisited. In *Proceedings of the 17th International Conference on The Theory and Application of Cryptology and Information Security* (Seoul, South Korea) *(ASIACRYPT'11)*. Springer-Verlag, Berlin, Heidelberg, 590–609. https://doi.org/10.1007/978-3-

[6] R.W. Baldwin. 1990. Naming and grouping privileges to simplify security management in large databases. In *Proceedings. 1990 IEEE Computer Society Symposium on Research in Security and Privacy*. 116–132. https://doi.org/10.1109/RISP.1990.63844

[7] Michael Ben-Or, Ran Canetti, and Oded Goldreich. 1993. Asynchronous secure computation. In *Proceedings of the Twenty-Fifth Annual ACM Symposium on Theory of Computing* (San Diego, California, USA) *(STOC '93)*. Association for Computing Machinery, New York, NY, USA, 52–61. https://doi.org/10.1145/167088.167109

[8] Michael Ben-Or, Ran Canetti, and Oded Goldreich. 1993. Asynchronous secure computation. In *Proceedings of the Twenty-Fifth Annual ACM Symposium on Theory of Computing* (San Diego, California, USA) *(STOC '93)*. Association for Computing Machinery, New York, NY, USA, 52–61. https://doi.org/10.1145/167088.167109

[9] Michael Ben-Or and Ran El-Yaniv. 2003. Resilient-optimal interactive consistency in constant time. *Distrib. Comput.* 16, 4 (Dec. 2003), 249–262. https://doi.org/10.1007/s00446-002-0083-3

[10] Michael Ben-Or, Shafi Goldwasser, and Avi Wigderson. 1988. Completeness theorems for non-cryptographic fault-tolerant distributed computation. In *Proceedings of the Twentieth Annual ACM Symposium on Theory of Computing* (Chicago, Illinois, USA) *(STOC '88)*. Association for Computing Machinery, New York, NY, USA, 1–10. https://doi.org/10.1145/62212.62213

[11] Michael Ben-Or, Boaz Kelmer, and Tal Rabin. 1994. Asynchronous secure computations with optimal resilience (extended abstract). In *Proceedings of the Thirteenth Annual ACM Symposium on Principles of Distributed Computing* (Los Angeles, California, USA) *(PODC '94)*. Association for Computing Machinery, New York, NY, USA, 183–192. https://doi.org/10.1145/197917.198088

[12] Josh Benaloh. 1987. *Verifiable Secret-Ballot Elections*. Ph. D. Dissertation. https://www.microsoft.com/en-us/research/publication/verifiable-secret-ballot-elections/

[13] Josh Benaloh, Ronald L. Rivest, Peter Y. A. Ryan, Philip B. Stark, Vanessa Teague, and Poorvi L. Vora. 2015. End-to-end verifiability. *CoRR* abs/1504.03778 (2015). arXiv:1504.03778 http://arxiv.org/abs/1504.03778

[14] David Bernhard, Véronique Cortier, David Galindo, Olivier Pereira, and Bogdan Warinschi. 2015. SoK: A Comprehensive Analysis of Game-Based Ballot Privacy Definitions. In *2015 IEEE Symposium on Security and Privacy*. 499–516. https://doi.org/10.1109/SP.2015.37

[15] G. R. BLAKLEY. 1979. Safeguarding cryptographic keys. In *1979 International Workshop on Managing Requirements Knowledge (MARK)*. 313–318. https://doi.org/10.1109/MARK.1979.8817296

[16] Daan Bloembergen, Davide Grossi, and Martin Lackner. 2019. On rational delegations in liquid democracy. In *Proceedings of the Thirty-Third AAAI Conference on Artificial Intelligence and Thirty-First Innovative Applications of Artificial Intelligence Conference and Ninth AAAI Symposium on Educational Advances in Artificial Intelligence* (Honolulu, Hawaii, USA) *(AAAI'19/IAAI'19/EAAI'19)*. AAAI Press, Article 221, 8 pages. https://doi.org/10.1609/aaai.v33i01.33011796

[17] Christian Blum and Christina Isabel Zuber. 2016. Liquid democracy: Potentials, problems, and perspectives. *Journal of political philosophy* 24, 2 (2016), 162–182.

[18] Gabriel Bracha. 1987. Asynchronous Byzantine agreement protocols. *Information and Computation* 75, 2 (1987), 130–143. https://doi.org/10.1016/0890-5401(87)90054-X

[19] D.F.C. Brewer and M.J. Nash. 1989. The Chinese Wall security policy. In *Proceedings. 1989 IEEE Symposium on Security and Privacy*. 206–214. https://doi.org/10.1109/SECPRI.1989.36295

[20] Christian Cachin, Rachid Guerraoui, and Lus Rodrigues. 2011. *Introduction to Reliable and Secure Distributed Programming* (2nd ed.). Springer Publishing Company, Incorporated.

[21] Christian Cachin, Klaus Kursawe, Anna Lysyanskaya, and Reto Strobl. 2002. Asynchronous verifiable secret sharing and proactive cryptosystems. In *Proceedings of the 9th ACM Conference on Computer and Communications Security* (Washington, DC, USA) *(CCS '02)*. Association for Computing Machinery, New York, NY, USA, 88–97. https://doi.org/10.1145/586110.586124

[22] Christian Cachin, Klaus Kursawe, and Victor Shoup. 2000. Random oracles in constantipole: practical asynchronous Byzantine agreement using cryptography (extended abstract). In *Proceedings of the Nineteenth Annual ACM Symposium on Principles of Distributed Computing* (Portland, Oregon, USA) *(PODC '00)*. Association for Computing Machinery, New York, NY, USA, 123–132. https://doi.org/10.1145/343477.343531

[23] Joseph Campbell, Alessandra Casella, Lucas de Lara, Victoria R. Mooers, and Dilip Ravindran. 2022. *Liquid Democracy. Two Experiments on Delegation in Voting*. NBER Working Papers 30794. National Bureau of Economic Research, Inc. https://doi.org/None

[24] Ran Canetti and Tal Rabin. 1993. Fast asynchronous Byzantine agreement with optimal resilience. In *Proceedings of the Twenty-Fifth Annual ACM Symposium on Theory of Computing* (San Diego, California, USA) *(STOC '93)*. Association for Computing Machinery, New York, NY, USA, 42–51. https://doi.org/10.1145/167088.167105

[25] Ioannis Caragiannis and Evi Micha. 2019. A contribution to the critique of liquid democracy. In *Proceedings of the 28th International Joint Conference on Artificial Intelligence* (Macao, China) *(IJCAI'19)*. AAAI Press, 116–122.

[26] David Chaum. 1983. Blind Signatures for Untraceable Payments. In *Advances in Cryptology*, David Chaum, Ronald L. Rivest, and Alan T. Sherman (Eds.). Springer US, Boston, MA, 199–203.

[27] David Chaum and Torben Pryds Pedersen. 1993. Wallet Databases with Observers. In *Advances in Cryptology — CRYPTO' 92*, Ernest F. Brickell (Ed.). Springer Berlin Heidelberg, Berlin, Heidelberg, 89–105.

[28] David L. Chaum. 1981. Untraceable electronic mail, return addresses, and digital pseudonyms. *Commun. ACM* 24, 2 (Feb. 1981), 84–90. https://doi.org/10.1145/358549.358563

[29] Benny Chor, Shafi Goldwasser, Silvio Micali, and Baruch Awerbuch. 1985. Verifiable secret sharing and achieving simultaneity in the presence of faults. In *26th Annual Symposium on Foundations of Computer Science (sfcs 1985)*. 383–395. https://doi.org/10.1109/SFCS.1985.64

[30] Zoé Christoff and Davide Grossi. 2017. Binary Voting with Delegable Proxy: An Analysis of Liquid Democracy. In *Theoretical Aspects of Rationality and Knowledge*. https://api.semanticscholar.org/CorpusID:7451305

[31] Véronique Cortier, David Galindo, Ralf Küsters, Johannes Müller, and Tomasz Truderung. 2016. SoK: Verifiability Notions for E-Voting Protocols. In *2016 IEEE Symposium on Security and Privacy (SP)*. 779–798. https://doi.org/10.1109/SP.2016.52

[32] Ronald Cramer, Ivan Damgård, and Berry Schoenmakers. 1994. Proofs of Partial Knowledge and Simplified Design of Witness Hiding Protocols. In *Advances in Cryptology — CRYPTO '94*, Yvo G. Desmedt (Ed.). Springer Berlin Heidelberg, Berlin, Heidelberg, 174–187.

[33] Ronald Cramer, Matthew Franklin, Berry Schoenmakers, and Moti Yung. 1996. Multi-Authority Secret-Ballot Elections with Linear Work. In *Advances in Cryptology — EUROCRYPT '96*, Ueli Maurer (Ed.). Springer Berlin Heidelberg, Berlin, Heidelberg, 72–83.

[34] Ivan Damgård, Mads Jurik, and Jesper Buus Nielsen. 2010. A generalization of Paillier's public-key system with applications to electronic voting. *Int. J. Inf. Secur.* 9, 6 (Dec. 2010), 371–385. https://doi.org/10.1007/s10207-010-0119-9

[35] Henry de Valence, Jack Grigg, Mike Hamburg, Isis Lovecruft, George Tankersley, and Filippo Valsorda. 2023. The ristretto255 and decaf448 Groups. RFC 9496. https://doi.org/10.17487/RFC9496

[36] Denise Demirel, Jeroen Van De Graaf, and Roberto Araújo. 2012. Improving Helios with everlasting privacy towards the public. In *Proceedings of the 2012 International Conference on Electronic Voting Technology/Workshop on Trustworthy Elections* (Bellevue, WA) *(EVT/WOTE'12)*. USENIX Association, USA, 8.

[37] Sisi Duan, Xin Wang, and Haibin Zhang. 2023. FIN: Practical Signature-Free Asynchronous Common Subset in Constant Time. In *Proceedings of the 2023 ACM SIGSAC Conference on Computer and Communications Security* (Copenhagen, Denmark) *(CCS '23)*. Association for Computing Machinery, New York, NY, USA, 815–829. https://doi.org/10.1145/3576915.3616633

[38] Cynthia Dwork, Nancy Lynch, and Larry Stockmeyer. 1988. Consensus in the presence of partial synchrony. *J. ACM* 35, 2 (April 1988), 288–323. https://doi.org/10.1145/42282.42283

[39] Paul Feldman. 1987. A practical scheme for non-interactive verifiable secret sharing. In *28th Annual Symposium on Foundations of Computer Science (sfcs 1987)*. 427–438. https://doi.org/10.1109/SFCS.1987.4

[40] Amos Fiat and Adi Shamir. 1987. How to prove yourself: practical solutions to identification and signature problems. In *Proceedings on Advances in Cryptology—CRYPTO '86* (Santa Barbara, California, USA). Springer-Verlag, Berlin, Heidelberg, 186–194.

[41] Kevin Gallagher, Santiago Torres-Arias, Nasir Memon, and Jessica Feldman. 2022. COLBAC: Shifting Cybersecurity from Hierarchical to Horizontal Designs. In *Proceedings of the 2021 New Security Paradigms Workshop* (Virtual Event, USA) *(NSPW '21)*. Association for Computing Machinery, New York, NY, USA, 13–27. https://doi.org/10.1145/3498891.3498903

[42] Huangyi Ge, Sze Yiu Chau, Victor E Gonsalves, Huian Li, Tianhao Wang, Xukai Zou, and Ninghui Li. 2019. Koinonia: verifiable e-voting with long-term privacy. In *Proceedings of the 35th Annual Computer Security Applications Conference* (San Juan, Puerto Rico, USA) *(ACSAC '19)*. Association for Computing Machinery, New York, NY, USA, 270–285. https://doi.org/10.1145/3359789.3359804

[43] S Goldwasser, S Micali, and C Rackoff. 1985. The knowledge complexity of interactive proof-systems. In *Proceedings of the Seventeenth Annual ACM Symposium on Theory of Computing* (Providence, Rhode Island, USA) *(STOC '85)*. Association for Computing Machinery, New York, NY, USA, 291–304. https://doi.org/10.1145/22145.22178

[44] Thomas Haines and Xavier Boyen. 2016. VOTOR: conceptually simple remote voting against tiny tyrants. In *Proceedings of the Australasian Computer Science Week Multiconference* (Canberra, Australia) *(ACSW '16)*. Association for Computing Machinery, New York, NY, USA, Article 32, 13 pages. https://doi.org/10.1145/2843043.2843362

[45] Thomas Haines, Rafieh Mosaheb, Johannes Müller, and Ivan Pryvalov. 2023. SoK: Secure E-Voting with Everlasting Privacy. *Proc. Priv. Enhancing Technol.* 2023 (2023), 279–293. https://api.semanticscholar.org/CorpusID:255921010

[46] Anson Kahng, Simon Mackenzie, and Ariel Procaccia. 2021. Liquid Democracy: An Algorithmic Perspective. *J. Artif. Int. Res.* 70 (May 2021), 1223–1252. https://doi.org/10.1613/jair.1.12261

[47] Paul A. Karger. 1987. Limiting the Damage Potential of Discretionary Trojan Horses. In *1987 IEEE Symposium on Security and Privacy*. 32–32. https://doi.org/10.1109/SP.1987.10011

[48] Christian Killer, Markus Knecht, Claude Müller, Bruno Rodrigues, Eder Scheid, Muriel Franco, and Burkhard Stiller. 2021. Æternum: A Decentralized Voting System with Unconditional Privacy. In *2021 IEEE International Conference on Blockchain and Cryptocurrency (ICBC)*. 1–9. https://doi.org/10.1109/ICBC51069.2021.9461101

[49] Christoph Carl Kling, Jerome Kunegis, Heinrich Hartmann, Markus Strohmaier, and Steffen Staab. 2015. Voting Behaviour and Power in Online Democracy: A Study of LiquidFeedback in Germany's Pirate Party. arXiv:1503.07723 [cs.CY] https://arxiv.org/abs/1503.07723

[50] Filip Kostelka. 2025. Understanding Voter Fatigue: Election Frequency and Electoral Abstention Approval. *British Journal of Political Science* 55 (2025), e85. https://doi.org/10.1017/S0007123425000171

[51] Jay Kreps. 2011. Kafka : a Distributed Messaging System for Log Processing. https://api.semanticscholar.org/CorpusID:18534081

[52] Leslie Lamport, Robert Shostak, and Marshall Pease. 1982. The Byzantine Generals Problem. *ACM Trans. Program. Lang. Syst.* 4, 3 (July 1982), 382–401. https://doi.org/10.1145/357172.357176

[53] Eric Landquist, Nimit Sawhney, and Simer Sawhney. 2025. Overcoming Bottlenecks in Homomorphic Encryption for the 2024 Mexican Federal Election. arXiv:2504.13198 [cs.CR] https://arxiv.org/abs/2504.13198

[54] Jure Leskovec, Daniel Huttenlocher, and Jon Kleinberg. 2010. Predicting positive and negative links in online social networks. In *Proceedings of the 19th international conference on World wide web*. 641–650.

[55] Jure Leskovec, Daniel Huttenlocher, and Jon Kleinberg. 2010. Signed networks in social media. In *Proceedings of the SIGCHI conference on human factors in computing systems*. 1361–1370.

[56] Maria Margarida Espanhol Lopes. 2023. *Mitigating rogue administrator attacks in democratic institutions using new access control methods*. Master's thesis.

[57] Achour Mostéfaoui, Hamouma Moumen, and Michel Raynal. 2015. Signature-Free Asynchronous Binary Byzantine Consensus with t < n/3, O(n2) Messages, and O(1) Expected Time. *J. ACM* 62, 4, Article 31 (Sept. 2015), 21 pages. https://doi.org/10.1145/2785953

[58] Open Web Application Security Project (OWASP). 2025. OWASP Top Ten:2025. https://owasp.org/Top10/2025/0x00_2025-Introduction/. Version 2025 of the OWASP Top Ten awareness document.

[59] Pascal Paillier. 1999. Public-Key Cryptosystems Based on Composite Degree Residuosity Classes. In *Advances in Cryptology — EUROCRYPT '99*, Jacques Stern (Ed.). Springer Berlin Heidelberg, Berlin, Heidelberg, 223–238.

[60] Torben P. Pedersen. 1991. Non-Interactive and Information-Theoretic Secure Verifiable Secret Sharing. In *Proceedings of the 11th Annual International Cryptology Conference on Advances in Cryptology (CRYPTO '91)*. Springer-Verlag, Berlin, Heidelberg, 129–140.

[61] Shaya Potter, Steven M. Bellovin, and Jason Nieh. 2009. Two-Person Control Administration: Preventing Administration Faults through Duplication. In *23rd Large Installation System Administration Conference (LISA 09)*. USENIX Association, Baltimore, MD. https://www.usenix.org/conference/lisa-09/two-person-control-administration-preventing-administration-faults-through

[62] Mark E. Russinovich and David A. Solomon. 2004. *Microsoft Windows Internals, Fourth Edition: Microsoft Windows Server(TM) 2003, Windows XP, and Windows 2000 (Pro-Developer)*. Microsoft Press, USA.

[63] Pierangela Samarati and Sabrina Capitani de Vimercati. 2001. Access Control: Policies, Models, and Mechanisms. In *Foundations of Security Analysis and Design*, Riccardo Focardi and Roberto Gorrieri (Eds.). Springer Berlin Heidelberg, Berlin, Heidelberg, 137–196.

[64] Ravi S Sandhu. 1998. Role-based access control. In *Advances in computers*. Vol. 46. Elsevier, 237–286.

[65] Adi Shamir. 1979. How to share a secret. *Commun. ACM* 22, 11 (Nov. 1979), 612–613. https://doi.org/10.1145/359168.359176

[66] Nathan Swearingen, Xukai Zou, and Ninghui Li. 2025. Fully Transparent, Privacy-Preserving Yet Verifiable, Attack-Resistant, and Practical Remote Electronic Voting Rendering Assured and Fair Elections. *IEEE Transactions on Privacy* 2 (2025), 105–118. https://doi.org/10.1109/TP.2025.3603141

[67] Latanya Sweeney. 2002. k-anonymity: A model for protecting privacy. *International journal of uncertainty, fuzziness and knowledge-based systems* 10, 05 (2002), 557–570.

[68] Jeroen van de Graaf. 2010. *Anonymous One-Time Broadcast Using Non-interactive Dining Cryptographer Nets with Applications to Voting*. Springer Berlin Heidelberg, Berlin, Heidelberg, 231–241. https://doi.org/10.1007/978-3-642-12980-3_14

[69] Ronghao Zhou and Zijing Lin. 2022. An Improved Exponential ElGamal Encryption Scheme with Additive Homomorphism. In *2022 International Conference on Blockchain Technology and Information Security (ICBCTIS)*. 25–27. https://doi.org/10.1109/ICBCTIS55569.2022.00017

## A   CPU Usage

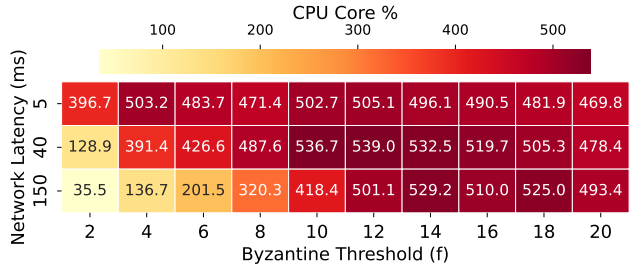

**Figure 6.** Average CPU usage

Figure 6 shows a heatmap with the average CPU usage of Voter instances as the induced network latency and number of processes increases. As with figure 3, the total number of Voter instances used is given by $3f + 1$, where $f$ is the byzantine threshold. We established in table 2 that each Voter has access to a set of 8 hardware cores, with hyperthreading enabled. As such, the maximum CPU usage achievable is 800% of a CPU core.

We observe that for a low byzantine threshold, the network latency has a significant impact on the CPU usage of the system. This is expected given that a communication bound process spends most of the execution's wall time idle, and is the most prominent with $f = 1$ and a network latency of $150\,ms$. As the byzantine threshold increases, the processes become computation bound regardless of the network latency induced. These observations corroborate the results from figure 3, where the induced network latency dominates the end-to-end latency of the e-voting procedure for low byzantine thresholds but fail to be noticeable for high thresholds.

## B   Emergency Votes Dual Threshold

During the voting procedure described in algorithm 3, the ABA instances act as a synchronization barrier: they force each process to either receive all vote shares or wait until a timeout occurs. Even if the timeout expires, the protocol blocks until at least $n - f$ vote shares have been received by correct processes.

Emergency votes (§ 5.3) introduce an exception to this execution flow by permitting an early termination as soon as $t_e$ vote shares are collected. In this setting, the protocol aims to produce an output as quickly as possible and therefore does not rely on timeouts. Algorithm ?? presents the extended Voter logic used during an Emergency vote. Under this logic, each Voter may produce at most two outputs: *EARLY* or *LATE*. A *LATE* output is always generated once the Voters agree on at least $n - f$ shares to be tallied. An *EARLY* output is optional and is produced only if the optimistic threshold $t_e$ is reached before the *LATE* output is triggered. While the protocol does not guarantee that the condition for an *EARLY* output will hold, the same argument used for proving the Termination property of the regular voting protocol ensures that every correct process eventually receives at least $n - f$ vote shares, and thus reaches a positive decision in at least $n - f$ ABA instances.

Because Voters must reach agreement on the tallies for both the *EARLY* and *LATE* outputs, two sets of ABA instances are required. Fortunately, a positive result from an *EARLY* ABA instance can be used to subsume agreement in the corresponding *LATE* instance, substantially reducing the overhead that would otherwise result from running two full sets of algorithms.

## C   Configuration

Privocracy supports configurable election thresholds, per-party voting weights, a *trust graph* for vote delegation, a maximum delegation weight threshold, such that no voter can surpass this threshold through vote delegation, and the elections' timeout. The parameters' explanation and example values are presented in table 3.

These configuration values are stored in each Privocracy daemon instance, and each daemon can have different configurations stored. For any given operation, as observed in algorithm 1, the specific configurations are loaded through the *loadConfig* command. This command calls an application specific script to load a file containing all the configurations. The configuration files are protected and can therefore only be created or modified through the use of the Privocracy command; allowing the introduction of new configuration values only if the majority of the voters agree on the introduced changes.

Table 3. Configurable parameters in Privocracy

| Parameter | Explanation | Example |
| --- | --- | --- |
| `weights` | Maps voters to their weights in the election | [(voter1, 5), (voter2, 3)...] |
| `threshold` | Percentage of votes, in cumulative weight, required to accept a command | 0.5 |
| `delegation` | Graph with the delegation relations between Voters | [(voter1, voter2, 5), (voter3, voter1, 2),...] |
| `maxWeight` | Maximum weight a Voter can have delegated, as a percentage of the total voters' weight | 0.3 |
| `timeout` | Timeout after which the voting algorithm stops waiting for all votes | 300 seconds |